# Possible spin-polarized Cooper pairing in high temperature FeSe superconductor

Yi Hu[1], Fanyu Meng[2], Hechang Lei[2], Qi-Kun Xue[1,3,4,5]* and Ding Zhang[1,3,5,6]*

[1] State Key Laboratory of Low Dimensional Quantum Physics and Department of Physics, Tsinghua University, Beijing 100084, China

[2] Department of Physics and Beijing Key Laboratory of Opto-electronic Functional Materials & Micro-nano Devices, Renmin University of China, Beijing 100872, China

[3] Beijing Academy of Quantum Information Sciences, Beijing 100193, China

[4] Southern University of Science and Technology, Shenzhen 518055, China

[5] Frontier Science Center for Quantum Information, Beijing 100084, China

[6] RIKEN Center for Emergent Matter Science (CEMS), Wako, Saitama 351-0198, Japan

*Email: qkxue@mail.tsinghua.edu.cn

dingzhang@mail.tsinghua.edu.cn

# Abstract

Superconductivity and long-range ferromagnetism hardly coexist in a uniform manner. The counter-example has been observed, in uranium-based superconductors for instance, with a coexisting temperature limited to about 1 K. Here, we report the coexistence of high temperature superconductivity and itinerant ferromagnetism in lithium intercalated FeSe flakes. In superconducting samples with transition temperature around 40 K, we observe the anomalous Hall effect with a hysteresis loop in transverse resistivity and a butterfly-like pattern of magneto-resistance. Intriguingly, such ferromagnetism persists down to a temperature at which the zero-field resistance fully vanishes. Furthermore, the superconductivity is enhanced under an in-plane magnetic field, suggestive of the participation of spin-polarized Cooper pairs. The surprising finding underscores a uniform coexistence of the two antagonistic phenomena on a record-high energy scale.

Combining superconductivity and long-range ferromagnetism is not only of fundamental interest in extending our understanding on distinct pairing mechanisms but also promises groundbreaking applications by interfacing dissipationless electronics with spintronics [1-3]. However, superconductivity occurs predominantly with singlet Cooper pairing and is incompatible with ferromagnetism that requires spins to be polarized to the same direction. Harmonizing superconductivity and ferromagnetism in the same material therefore requires ingenuity and serendipity. Several U-based superconductors—UGe$_2$ [4], URhGe [5], UCoGe [6] and lately UTe$_2$ [7,8]—have been reported to show strong and uniform interaction between Cooper pairing and ferromagnetism. These superconducting ferromagnets are considered as candidates for the long sought-after triplet pairing. In these systems, the superconducting transition temperature $T_{sc}$ only lingers around 1 K and the Curie temperature $T_m$ stays below 30 K. One possible route to boost $T_{sc}$ and $T_m$ is by stacking cuprate or iron-based superconductors with ferromagnetic layers [3,9-13]. It takes advantage of the layered structure of the high-$T_{sc}$ superconductors. However, the coupling between superconductivity and ferromagnetism is usually weak due to the large spatial separation (in the *c*-axis) and often distinct orbitals that account for itinerant electrons and magnetic moments.

Here, we report uniform coexistence of superconductivity and itinerant ferromagnetism at high $T_{sc}$ of 40 K and $T_m > 200$ K in lithium intercalated FeSe flakes. The coexistence is further supported by enhanced superconductivity under an in-plane magnetic field. Our results have profound implications on the pairing mechanism of iron-based superconductors and establish a distinct route that combines high-$T_{sc}$ superconducting electronics and high-$T_m$ spintronics.

We carry out lithium intercalation into FeSe via a solid-state ion backgate. This technique has been intensively employed in tuning FeSe before [14-16] and applied to a slew of other materials too [17-24]. Figure 1a schematically illustrates our samples. An exfoliated flake of FeSe capped by the hexagonal boron nitride (h-BN) is positioned on the solid ion conductors (SIC) and contacted by the bottom electrodes (details in Methods). We apply a positive backgate voltage at room temperature (300 K) to intercalate lithium ions into FeSe and terminate the ion motion by cooling to low temperatures (typically below 200 K). The uniformity of lithium distribution in the c-axis is guaranteed by the relatively small thickness of the flakes, as confirmed in previous studies [14-16]. Figure 1b reveals the enhanced superconductivity. We hereupon focus on sample S1 (14 nm thick) for the transport results. Lithium intercalation promotes $T_{sc}$ to a value exceeding 40 K, in agreement with previous studies. We collect data from different pairs of contacts on the same sample (as shown in the inset of Fig. 1b). Except for the pair 4-5 (upper left region of the sample in the inset), other three pairs of contacts show the same superconducting transition temperature $T_{sc,0}$ =42 K (We define $T_{sc,0}$ as the temperature point above which the zero-field resistance reaches 1% of the normal state resistance). The overlapping behavior indicates a uniform distribution of lithium in the *a-b* plane of FeSe.

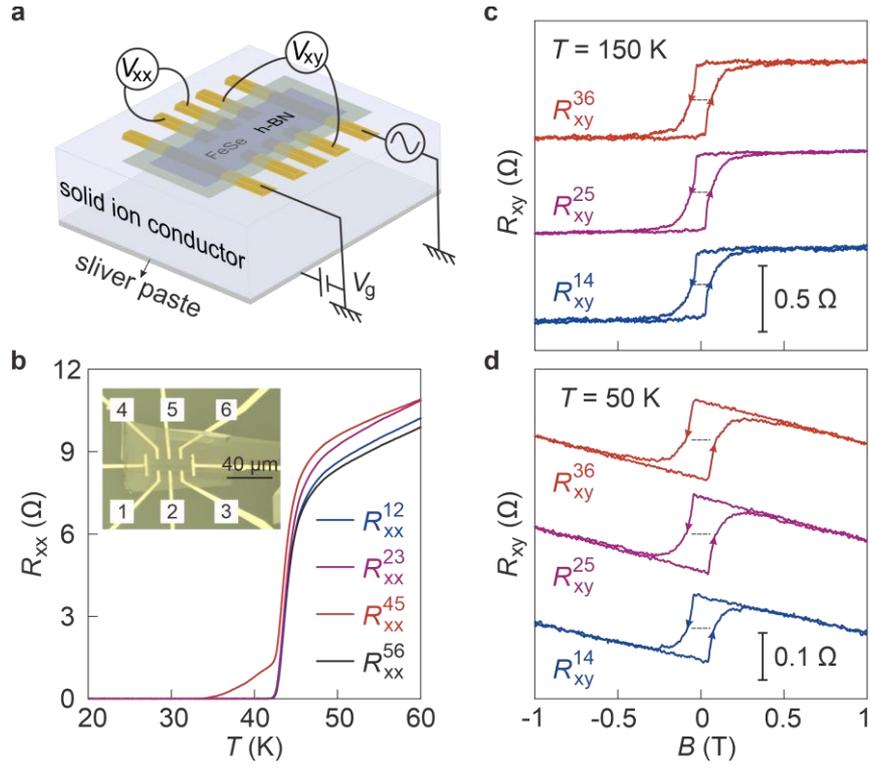

**Figure 1 Coexistence of superconductivity and anomalous Hall effect in lithium intercalated FeSe. a** Schematic drawing of the experimental setup. **b** Temperature dependent resistances measured from different pairs of contacts on sample S1 after lithium intercalation. Inset is an optical image of sample S1, taken before capping with h-BN. Upper indices of $R_{xx}$ correspond to the contacts shown in the inset. **c** and **d** Hall resistances as a function of perpendicular magnetic field from three pairs of contacts on the same FeSe sample after lithium intercalation. Upper indices of $R_{xy}$ represent the contacts used (shown in the inset of Fig. 2b). They are measured at 150 K (**c**) and 50 K (**d**). Data are anti-symmetrized from the raw data (see Methods). Curves are vertically offset for clarity. Dashed lines mark zero Hall resistances for different data sets. Arrows indicate the sweeping direction of the magnetic field.

Figure 1c and d present evidence for ferromagnetism in our Li-intercalated FeSe. Hall resistances measured from three pairs of contacts—with a longitudinal separation of 10 μm between each pair—show the same hysteresis loop from the anomalous Hall effect (AHE), attesting to a rather uniform ferromagnetic state (The pristine sample only shows a linear Hall response). On the higher temperature side, the hysteretic behavior can also be seen at 180/200 K. Further increasing the temperature results in activation of the lithium intercalation/de-intercalation process such that the field sweep does not show a closed hysteresis loop. Nevertheless, it indicates that $T_m$ must be much higher than 200 K. On the lower temperature side, the ordinary Hall effect (OHE) at high magnetic fields changes sign of its slope at about 125 K. It indicates that n-type carriers dominate the transport at low temperatures, consistent with previous reports [14,16].

A similar observation of AHE in Li-intercalated FeSe at 50 K was speculated to arise from locally concentrated intercalates of lithium, which induce non-superconducting ferromagnetic puddles inside the high-$T_{sc}$ matrix of FeSe [16,13]. We point out that this scenario of phase separation seems unlikely here. First of all, we observe pronounced hysteresis of AHE in a superconducting state with a steep superconducting transition, indicating homogeneous doping distribution. Secondly, phase separation, on a mesoscopic scale, is incompatible with the observation of the same hysteresis loop at distinctly different positions of the sample. The uniform coexistence of superconductivity and ferromagnetism in our sample is further supported by the experimental findings given below.

Figure 2 shows a systematic evolution of the AHE (Fig. 2a, b, d) and the corresponding longitudinal resistance (Fig. 2c, e). As shown in Fig. 2a, the linear extrapolation of $R_{xy}$ from the high magnetic field to $B = 0$ T (dashed lines)—$R_{AH}$—is positive but flips to a negative value at temperatures below 45 K. This sign reversal of AHE suggests the existence of two competing anomalous Hall components. We note that the OHE shows little variation throughout this transition. Figure 2f further provides quantitatively the extracted $R_{AH}$ (diamonds) and Hall density (circles) as a function of temperature. The nearly constant density in the superconducting transition regime (double arrows) corresponds to a fixed Fermi level thus little change in the Berry curvature. Therefore, the anomalous Hall component from the non-trivial Berry curvature [25,26] should stay unchanged throughout the superconducting transition. Apart from this intrinsic mechanism, skew scattering can contribute to the AHE [25]. Indeed, $R_{AH}$ scales linearly with $R_{xx}$ for the data in the temperature window from 180 to 50 K, supporting the presence of skew scattering [25]. These scattering processes may give rise to a positive anomalous Hall signal while the intrinsic mechanism has a negative contribution. By further lowering the temperature from 50 K, the condensation of Cooper pairs dramatically suppresses scattering, leading to a sign reversal of $R_{AH}$. We note that vortex dynamics around $T_{sc,0}$ may produce a non-linear Hall signal [27,28] but it typically hosts an opposite sign than the OHE. This is different from our results that both AHE and OHE show negative slopes at $T < 43$ K. At temperatures below 40 K (Fig. 2d), the AHE gets strongly suppressed and the transverse resistance becomes linear with the magnetic field after exceeding a certain critical field.

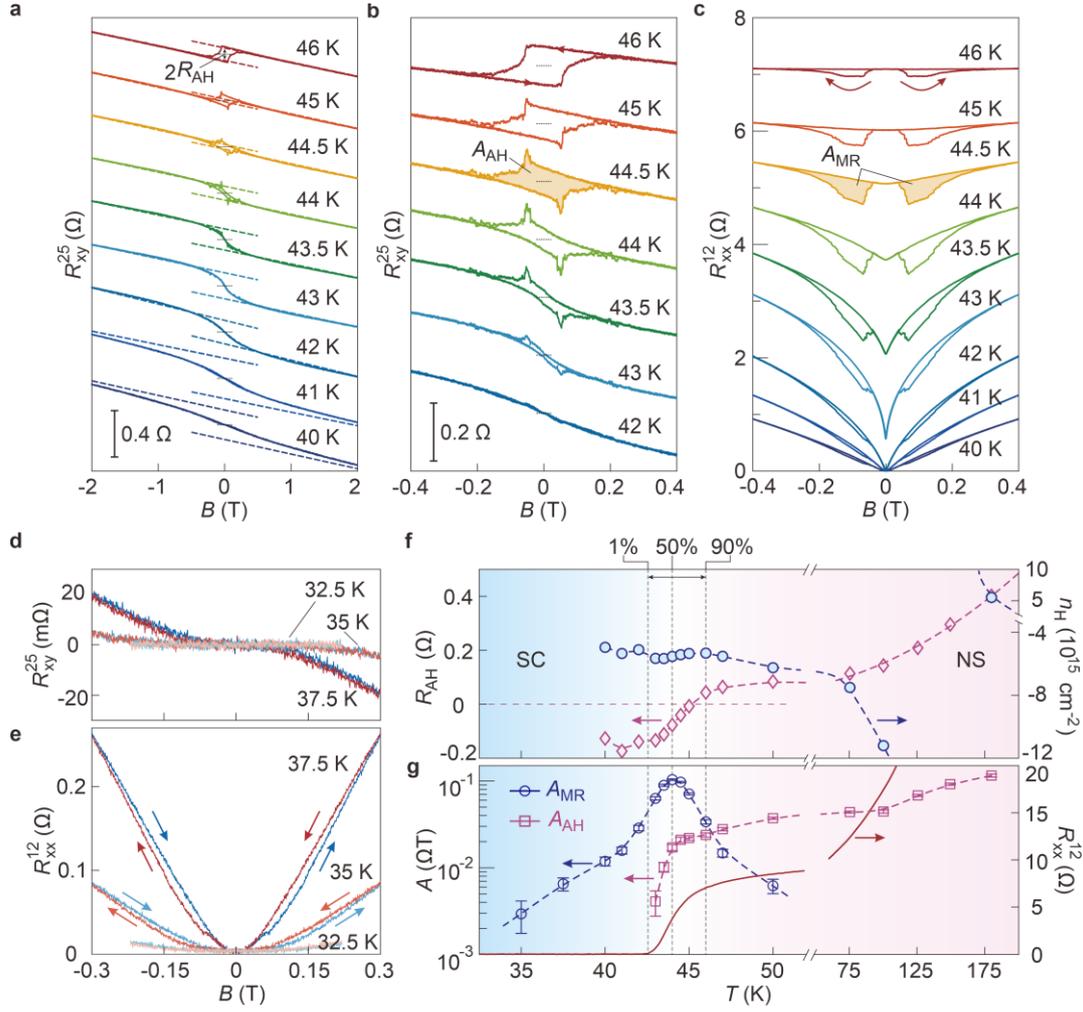

*Figure 2 Anomalous Hall effect and magnetoresistance in lithium intercalated FeSe.* Data are collected from sample S1. Upper indices of $R_{xy,xx}$ denote the pair of contacts used (inset of Fig. 1b). **a, b, d** Hall resistance as a function of perpendicular magnetic field measured at a set of temperature points. Curves are vertically offset in **a** and **b** (dotted lines indicate zero Hall resistances). Dashed lines in **a** are linear fits to the data at high fields. Black arrows in **a** indicate the definition of the anomalous Hall resistance $R_{AH}$. Shaded region in **b** illustrates the area of the hysteresis loop $A_{AH}$. **c, e** Longitudinal resistance as a function of perpendicular magnetic field at a set of temperature points. Shaded region in **c** illustrates the area of the hysteresis loop $A_{MR}$. Arrows denote the sweeping directions. **f** Temperature dependence of $R_{AH}$ (diamonds) and the Hall carrier density (circles). Double arrows indicate the superconducting transition regime. Vertical lines indicate the positions where the zero-field resistance reaches 1%, 50% and 90% of the normal state value. Color shades indicate the superconducting (SC) and normal state (NS) regimes. **g** Temperature dependence of the hysteresis loop areas $A_{AH}$ (squares) and $A_{MR}$ (circles). Error bars are obtained from the standard deviation of the corresponding field-sweeping data at high magnetic fields (0.3 to 0.4 T). Solid curve: temperature dependent resistance at zero field.

Interestingly, we observe a peak/dip structure in the Hall traces around the coercive field at 42 to 45 K in Fig. 2b, which is reminiscent to the topological Hall effect (THE) [29,30] and suggestive of non-trivial spin textures in a high-$T_{sc}$ superconductor [31]. However, such a peak-dip is absent in the other pair of the Hall probes (contacts 1 and 4 in inset of Fig. 1b), indicating its possible sensitivity to local inhomogeneity. This feature may also reflect the competition between two anomalous Hall components [32,33]. We remark that the general temperature evolution of the AHE and OHE stays the same from the other pair.

Accompanied with the transverse resistance, the longitudinal resistance $R_{xx}$ shows the typical butterfly pattern by sweeping the magnetic field back and forth (Fig. 2c, e). Even at 35 K ($83\%T_{sc,0}$), clear hysteresis can be observed (Fig. 2e). We exclude the magnetocaloric effect here by measuring the hysteresis at a sufficiently low sweeping rate. Figure 2g summarizes the encircled areas of hysteresis loops, as illustrated by the shaded areas in Fig. 2b and c, for the anomalous Hall and longitudinal resistance traces, respectively. In contrast to a monotonic suppression of the loop area in the anomalous Hall signal $A_{AH}$, the loop area in the longitudinal resistance $A_{MR}$ shows a non-monotonic evolution with decreasing temperature. From 46 to 44 K, i.e., from the superconducting onset temperature to the mid-point of the transition, $A_{MR}$ actually becomes more pronounced. This enhancement is possibly due to strongly randomized magnetic domains as superconducting puddles start to emerge. The loop area $A_{MR}$ becomes weaker at lower temperatures as superconductivity becomes stronger. Nevertheless, $A_{MR}$ stays observable even deep in the superconducting state, indicating profound interplay between superconductivity and ferromagnetism.

We further rotate the sample such that the orientation of the magnetic field changes from the out-of-plane direction ($\theta = 90°$) to the in-plane case ($\theta = 0°$) (inset of Fig. 3a). Figure 3a and 3b show the magnetic field dependences of Hall and longitudinal resistances in the normal state (50 K) at several selected tilting angles.

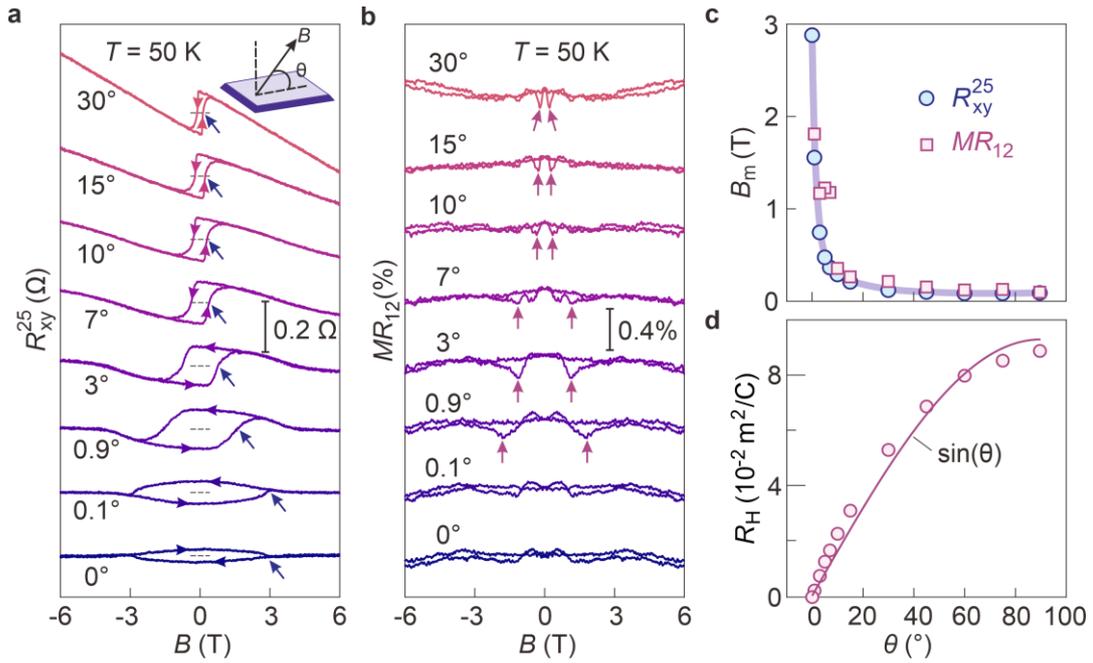

***Figure 3 Tilt-field study of the ferromagnetic state.*** *Data are all obtained from sample S1 at 50 K.* ***a*** *Hall resistance as a function of magnetic field at selected tilting angles. Inset illustrates the definition of the tilting angle. Curves are vertically offset with the dashed lines representing zero Hall resistances. Arrows on the curves indicate the sweeping directions. Dark blue arrows point out the positions of the coercive fields $B_m$.* ***b*** *Magnetoresistance [MR=$(R_{xx}(B) - R_{xx}(0))/R_{xx}(0)$] at selected tilting angles. Arrows mark the local minima in the curves. Their magnetic field positions are taken as the coercive fields.* ***c*** *Coercive fields from Hall and longitudinal resistances as a function of tilting angle.* ***d*** *Angular dependence of the Hall coefficients obtained from the high field section: [4, 8] T. Solid curve is a fit by using the $sin(\theta)$ dependence.*

With decreasing θ, the hysteresis loop in $R_{xy}(B)$ gets squeezed along the vertical axis and stretched in the horizontal axis. It indicates that the AHE becomes suppressed with the magnetic field tilting to the in-plane direction and the coercive field becomes larger. The latter feature is also confirmed by the magneto-resistance data in Fig. 3b (marked by arrows). These results are consistent with the scenario that the ferromagnetic easy-axis aligns with the *c*-axis of FeSe. In addition, we estimate a slight misalignment in θ of about 0.03 degree based on the sign-reversal in the chirality of the hysteresis loop when tilting from 0.1° to 0° (indicated by the arrows on the bottom two curves in Fig. 3a). In Fig. 3c, d. we summarize the angular dependence of the coercive fields (indicated by the arrows in Fig. 3a, b) and the Hall coefficient (estimated from the high field section). The Hall coefficient decreases continuously with the reduction of θ, following $sin(\theta)$ dependence (Fig. 3d). It indicates that the thin FeSe flake behaves predominantly as a two-dimensional electronic system.

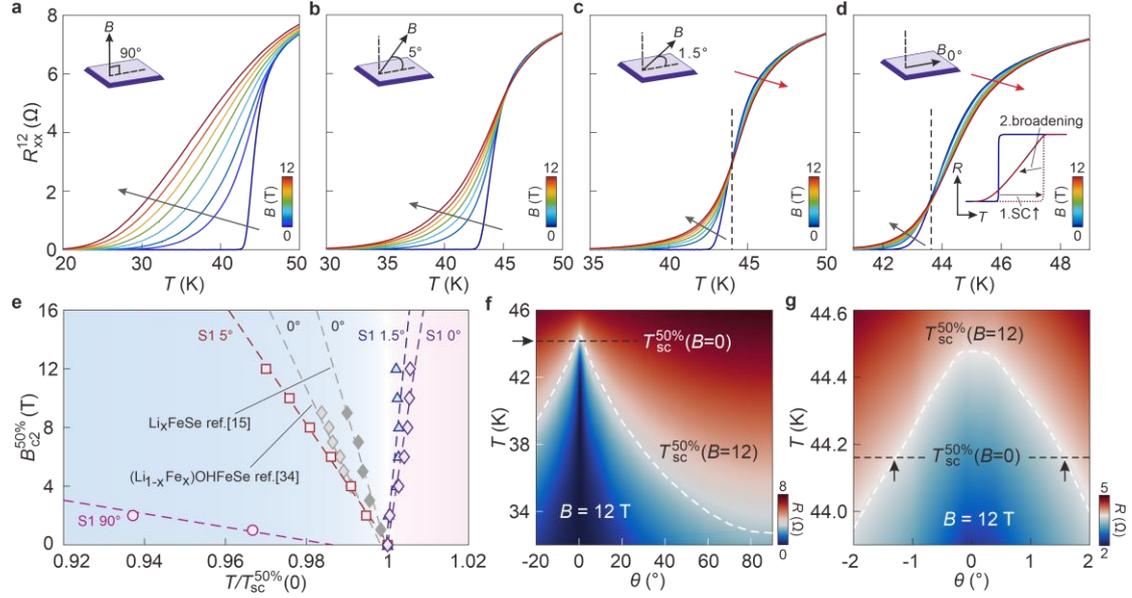

***Figure 4 Enhanced superconductivity under an in-plane magnetic field. a-d*** *Temperature dependent longitudinal resistances at a set of magnetic fields for sample S1. They are measured with different tilting angles as schematically illustrated in the insets. Gray and red arrows indicate the conventional and anomalous field responses, respectively. Inset in **d** illustrates two effects involved when applying an in-plane magnetic field: 1. Enhanced superconductivity (SC); 2. Increased broadening. **e** Upper critical fields as a function of normalized temperature at different tilting angles. Here we apply the mid-point criterion to data in panels **a-d**. The dark grey diamonds correspond to in-plane $B_{c2}$ for 44 K phase of $Li_xFeSe$ [15]. The light grey diamonds are obtained from $(Li_{1-x}Fe_x)OHFeSe$ [34]. **f, g** Color-plot of the longitudinal resistance at 12 T as a function of temperature and tilting angle. Black dashed line represents the superconducting transition temperature defined by the mid-point criterion at zero magnetic field: $T_{sc}^{50\%}(B=0)$. White dashed curve represents the transition temperature at 12 T: $T_{sc}^{50\%}(B=12)$. Arrows in **g** indicate the critical angles. Within this small range, $T_{sc}^{50\%}(B=12) > T_{sc}^{50\%}(B=0)$.*

We now unveil evidence of unconventional superconductivity due to the presence of ferromagnetism. Figure 4a to d show the superconducting transition in resistance at a set of magnetic fields under different tilting. Typically, applying a magnetic field weakens superconductivity such that the resistance traces at higher $B$ shift to lower temperatures. This standard behavior is confirmed in Fig. 4a, b but is obviously violated in Fig. 4c or d when the magnetic field is applied close to the in-plane direction. There, applying a larger magnetic field seems to shift a large portion of the curves to higher temperatures (indicated by the red arrows), although the section at lower temperatures seem to show the standard response (gray arrows). By taking the mid-point criterion ($50\% R_n$) to define the critical magnetic field, we summarize the contrasting temperature dependences in Fig. 4e. Whereas $B_{c2}$ vs. $T/T_{sc}^{50\%}(0)$ shows the conventional negative slope at the tilting angles of 90° and 5°, it changes to a positive slope when the angle is close to 0°. This behavior is distinctly different from the well-documented magnetic field responses of iron-based superconductors with comparable $T_{sc}$ [15,34]. Still, taking a criterion of $10\% R_n$ would generate a seemingly standard response. To understand such a dichotomy, we argue that two effects are at play in the superconductor under a high magnetic field: (1) shift of the transition to a higher temperature; (2) broadening of the transition. Together, they give rise to the crossing of the curves obtained at different magnetic fields, as schematically illustrated in Fig. 4d. While the second effect is trivial and stems probably from the dissipation of vortices, the first effect points to enhanced superconductivity under an in-plane magnetic field.

The sensitivity of the enhanced superconductivity to the tilting can be appreciated by measuring the temperature dependent resistance at various $\theta$ but a fixed magnetic field. Figure 4f shows a color-coded plot of the temperature dependent resistance at 12 T. The white dashed line demarcates the superconducting transition temperature at 12 T, i.e., $T_{sc}^{50\%}(B = 12)$. From this measurement, it becomes clear that $T_{sc}^{50\%}(B = 12)$ around $\theta \leq \pm 1.5°$ is higher than the zero-field value (Fig. 4g). Enhanced/re-entrant superconductivity under an in-plane magnetic field was previously reported in Pb films [35], LaAlO₃/SrTiO₃ [35], and twisted bilayer/double bilayer/trilayer graphene [36-38]—all with rather low $T_{sc}$. In the in-plane direction, the orbital pair-breaking effect of the magnetic field is largely suppressed. Instead, Zeeman splitting acts as the main mechanism for destroying singlet pairing. However, the spin polarization is favored in a ferromagnetic superconductor, giving rise to enhanced $T_{sc}$ with increasing magnetic field. In general, the observation of enhanced superconductivity with $T_{sc} \sim 40$ K under an in-plane magnetic field, together with strong evidence for a uniformly coexisting ferromagnetic order, suggest non-trivial Cooper pairing. Lithium intercalated FeSe is therefore a promising platform that interfaces superconductivity and spintronics for potential applications.

**METHODS**

**Sample growth and fabrication**

Single crystals of FeSe were grown by a temperature-gradient assisted flux method [39]. High-purity Fe and Se powder were mixed in a quartz ampoule at a 1:1 ratio with a eutectic mixture of $AlCl_3$ and NaCl in the ratio of 0.52:0.48. The ampoule was evacuated down to 0.01 mbar, sealed, and placed horizontally in a furnace with a uniform temperature gradient from 620 K to 700 K for 30 days. After the growth period, we switched off the furnace and let the ampoule cool down to room temperature. After rinsing multiple times with distilled water and ethanol, FeSe crystals were dried in an oven at 100°C for a few minutes and stored in a glovebox with Ar atmosphere.

The pristine FeSe crystals exhibited an onset superconducting transition temperature of 8 K. The Energy Dispersive X-ray (EDX) spectroscopy indicated a Fe:Se ratio of 0.92:1, suggesting a slight deficiency of iron. The crystals were mechanically exfoliated in an argon-glovebox ($H_2O$<0.1 ppm and $O_2$<0.1 ppm) [17]. The exfoliated flakes were dry transferred onto the solid ion conductors (chemical formula: $Li_2O$-$Al_2O_3$-$SiO_2$-$P_2O_5$-$TiO_2$-$GeO_2$, from Ohra corp.) with prepatterned electrodes of Ti and Au (10/30nm) by using electron beam lithography. We transferred a h-BN flake to cover FeSe for protection. Before transport measurements, the sample thickness was determined by atomic force microscopy (AFM) by scanning over a corner of the FeSe flake that was not covered by h-BN.

Former studies estimated the Li:Fe ratio based on the charging current during the lithium intercalation process. They further obtained $T_{sc}$ for samples with different Li:Fe ratios. From these studies, $T_{sc} \sim 40$ K, as measured from our samples, corresponds to the Li:Fe ratio of 0.2 [14] or 0.35 [15].

**Measurement**

The electrical transport measurements were carried out by using two closed-cycle cryogenic systems equipped with superconducting magnets. We employed the standard low-frequency lock-in technique with an excitation current of 5 µA (13.33 Hz). A DC source-meter was used to apply the back-gate voltage at 300 K. The resistance of sample S1 dropped quickly when the gate voltage was ramped to 3.2 V. We cooled down the sample when the resistance decreased to about 40% of its initial value.

For the sample rotation, we employed a home-built insert with a piezo-driven rotator (angular precision: 0.006°). The rotation axis is perpendicular to the axis of the solenoid magnet. For obtaining the data in Fig. 4d, we monitored the sample resistance and continuously rotated the sample at a fixed magnetic field of 2 T and a fixed temperature of 40 K (in the superconducting transition regime). The in-plane direction was chosen as the angle at which the resistance showed a minimum. It utilized the strong anisotropy of the upper critical magnetic fields in Li-intercalated FeSe.

The magnetic field sweep rate was 0.02 T/min for the data presented in Fig. 1c, 0.1 T/min for the data in Fig 1d and Fig. 3a,b, 0.05 T/min for the data in Fig. 2a, and 0.005 T/min for the data in Fig. 2b-e. We presented in the main text the magnetic field dependent Hall and longitudinal resistances after anti-symmetrizing/symmetrizing the raw data to exclude the possible mixing of the two diagonal resistance components. Specifically, we took two field sweeps in the opposite directions and measured $R_{H,\rightarrow}(B)$, $R_{H,\leftarrow}(B)$, $R_{l,\rightarrow}(B)$, and $R_{l,\leftarrow}(B)$. Here, $H$ and $l$ indicate Hall and longitudinal resistances. Right/left arrows indicate the sweeping directions. We

calculated $R_{xy}(B)$ and $R_{xx}(B)$ by using:

$$R_{xy}(B) = [R_{H,\rightarrow}(B) - R_{H,\leftarrow}(-B)]/2,$$

$$R_{xx}(B) = [R_{l,\rightarrow}(B) + R_{l,\leftarrow}(-B)]/2.$$

Both $R_{xy,xx}(B)$ and $R_{xy,xx}(-B)$ are shown in the figures. We remark that pronounced hysteresis already exists in the raw data.

**Acknowledgements**

We thank Denis Maryenko, Vadim Grinenko, Klaus von Klitzing for critical reading of the manuscript and insightful discussions. D. Z. and Q.-K. X. acknowledge financial support from the Ministry of Science and Technology of China (2022YFA1403100); D. Z. acknowledges financial support from National Natural Science Foundation of China (Grants No. 12361141820, No. 12274249, No. 52388201); H. C. L. acknowledges financial support from Beijing Natural Science Foundation (Grant No. Z200005).

**Author contributions**

D. Z. and Q.-K. X. designed the project. Y. H. fabricated the samples and did the transport measurements. Y. H. and F. M. grew the crystals with the guidance of H. C. L. D. Z., Y. H. analyzed the data with the input of Q.-K. X. D. Z. and Y. H. wrote the paper with the input of H. C. L. and Q.-K. X.